\documentstyle[epsfig,11pt]{article}
\newcommand{\zp}[3]{Z. Phys.\ C#1 (19#2) #3}
\newcommand{\pl}[3]{Phys.\ Lett.\ #1B (19#2) #3}
\newcommand{\np}[3]{Nucl.\ Phys.\ B#1 (19#2) #3}
\newcommand{\prd}[3]{Phys.\ Rev.\ D#1 (19#2) #3}
\newcommand{\prl}[3]{Phys.\ Rev.\ Lett.\ #1 (19#2) #3}

\newcommand{\gaga}{\mbox{$\gamma\gamma$}}

\def\simgt{\rlap{\lower 3.5 pt \hbox{$\mathchar \sim$}} \raise 1pt \hbox {$>$}}
\def\simlt{\rlap{\lower 3.5 pt \hbox{$\mathchar \sim$}} \raise 1pt \hbox {$<$}}

\newcommand{\epem}{\mbox{$e^+e^-$}}
\newcommand{\qqb}[1]{\mbox{$#1\overline{#1}$}}

\newcommand{\beq}{\begin{equation}}
\newcommand{\eeq}{\end{equation}}
\newcommand{\bea}{\begin{eqnarray}}
\newcommand{\eea}{\end{eqnarray}}

\catcode`\@=11

\def\@citex[#1]#2{\if@filesw\immediate\write\@auxout{\string\citation{#2}}\fi
  \def\@citea{}\@cite{\@for\@citeb:=#2\do
    {\@citea\def\@citea{,\penalty\@m}\@ifundefined
       {b@\@citeb}{{\bf ?}\@warning
       {Citation `\@citeb' on page \thepage \space undefined}}%
\hbox{\csname b@\@citeb\endcsname}}}{#1}}

\def\citer{\@ifnextchar [{\@tempswatrue\@citexr}{\@tempswafalse\@citexr[]}}

\def\@citexr[#1]#2{\if@filesw\immediate\write\@auxout{\string\citation{#2}}\fi
  \def\@citea{}\@cite{\@for\@citeb:=#2\do
    {\@citea\def\@citea{--\penalty\@m}\@ifundefined
       {b@\@citeb}{{\bf ?}\@warning
       {Citation `\@citeb' on page \thepage \space undefined}}%
\hbox{\csname b@\@citeb\endcsname}}}{#1}}
\relax

\begin{document}

\thispagestyle{empty}

\hfill\vbox{
                                   }
\vspace{1.0in}

\begin{center}
{\Large\bf Photoproduction of Heavy Quarks$\,^*$} \\

\vspace{1.5cm}

{\large \sc Michael~Kr\"amer}$\,^\dagger$

\vspace{0.3cm}

{\em Deutsches Elektronen-Synchrotron DESY, D-22603 Hamburg, Germany}

\end{center}

\vspace{1.5cm}

\noindent
\begin{abstract}
Heavy quarks are copiously produced in two-photon collisions at $e^+e^-$
colliders. The theoretical predictions including QCD radiative corrections
are compared to recent experimental data on $\gamma\gamma$ production of
charm quarks at PETRA, PEP, TRISTAN and LEP. Photoproduction of heavy quarks
at HERA is an important tool to measure the gluon distribution in the proton.
New theoretical results on heavy quark photoproduction at large transverse
momenta are discussed and NLO predictions for inelastic $J/\psi$
photoproduction in the HERA energy range are given. The sensitivity of the
results to the parametrization of the gluon distribution in the small-$x$
region is demonstrated.
\end{abstract}

\vfill

\noindent
{
$*\,$ Talk presented at the Conference "Photon '95", Sheffield, UK,
      April 1995;\\
      \hspace*{3.mm} to appear in the proceedings.\\
$\dagger\,$ E-mail: mkraemer@desy.de}
\newpage

\vspace{0.5cm}

{\noindent\large\bf I.~Introduction}\\[3mm]
The study of heavy flavour production is one of the important areas of
research at present and future high-energy colliders. Many features of the
production mechanism for heavy quarks are calculable in perturbative QCD.
The mass of the heavy quark, $m \gg \Lambda_{\mbox{\scriptsize QCD}}$, acts
as a cutoff and sets the scale for the perturbative calculations.
The production cross section factorizes into a partonic hard scattering
cross section multiplied by light quark and gluon parton densities
\cite{coll}. In order for this approach to be valid the mass of the produced
quarks must be sufficiently large for two reasons. First, contributions to
the cross section are neglected that are suppressed by powers of
$\Lambda_{\mbox{\scriptsize QCD}}/m$. After neglecting such power suppressed
contributions in order to arrive at a factorized form, one expands the
hard scattering cross section in powers of the strong coupling constant
$\alpha_s(m)$, evaluated at a scale set approximately by the heavy quark
mass. It is thus not guaranteed {\em a priori} that charm and bottom quark
production can reliably be predicted in perturbative QCD. Higher-twist
uncertainties of order $(\Lambda_{\mbox{\scriptsize QCD}}/m_c) \sim 20-30\%$
have to be taken into account. Moreover, the convergence of the perturbative
expansion might be poor due to the large value of the strong coupling
constant at the charm mass scale, $\alpha_s(m_c) \sim 0.3$. The situation is
much better for bottom production, although even there the theoretical
uncertainties can be large. The main questions therefore are whether
perturbative QCD calculations are trustworthy already for charm and bottom
quark production and whether the dynamics of these processes can reliably be
described without including non-perturbative effects.
The simplest process to study the validity of the perturbative approach
is heavy quark production in two-photon collisions. The predictions for the
dominant direct production channel do not depend on quark and gluon
distributions in the photon so that they are free of phenomenological
parameters. They depend only on the heavy quark mass and the QCD coupling.

\vspace{0.5cm}

{\noindent\large\bf
II.~Heavy-Quark Production in Two-Photon Collisions}\\[3mm]
Three mechanisms contribute to the production of heavy quarks in \gaga\
collisions: $(i)$ In the case of direct production, the photons couple
directly to the heavy quarks. No spectator particles travel along the
$\gamma$~axes. $(ii)$ If one of the photons first splits into a flux of light
quarks and gluons \cite{phsplit}, one of the gluons may fuse with the second
photon to form the \qqb{Q} pair. The remaining light quarks and gluons build
up a spectator jet in the split~$\gamma$ direction (single resolved $\gamma$
contribution). The total \gaga\ cross section of this mechanism depends on
the parton density of the photon \cite{photstruc}. Since the number of gluons
in the resolved photon grows $\sim\alpha\alpha_s^{-1}$, the resolved $\gamma$
processes are of the same order as the direct process. $(iii)$ If both
photons split into quarks and gluons, the \qqb{Q} pair is accompanied by two
spectator jets (double resolved $\gamma$ contribution). It turns out {\it a
posteriori} that the double resolved $\gamma$ contribution is much smaller
than the direct and the single resolved $\gamma$ contributions.

Two-photon production of heavy quarks can be studied at high-energy
\epem~col\-li\-ders where a large number of equivalent photons is
generated.\footnote{Heavy quark production in deep-inelastic
$e\gamma$ scattering has been discussed by E.~Laenen at this conference
\cite{eric}.}\,
Total cross sections and various distributions for charm and bottom
quark production $\epem\to\epem c/b X$ have been calculated in
Refs.~\cite{phd,dkzz} including QCD radiative corrections for the leading
subprocesses. The main results of that analysis can be summarized as follows:
At low energies in the PETRA/PEP/TRISTAN range, the direct production
mechanism by far dominates the total cross section. At LEP2~$\sim$~180~GeV,
the direct contribution and the 1-resolved $\gamma$ contribution are of equal
size. While the cross sections for charmed particle production are large,
giving a total of $\sim$~350,000 events for an integrated luminosity of
$\int{\cal L} = 500$~pb$^{-1}$ at LEP2, $b$~quark production is suppressed
by more than two orders of magnitude. The QCD corrections are important,
increasing the cross sections by $\sim$~30\%. At the \epem energies reached
up to now the theoretical expectations do not depend much on the
parametrization for the quark and gluon densities of the photon. At higher
energies the resolved processes are predicted to overtake the direct
contribution. A measurement of the gluon content of the photon, which is
currently poorly known, might thus be feasible at LEP2.

\begin{figure}[hbtp]

\hspace*{1.75cm}
\epsfig{%
file=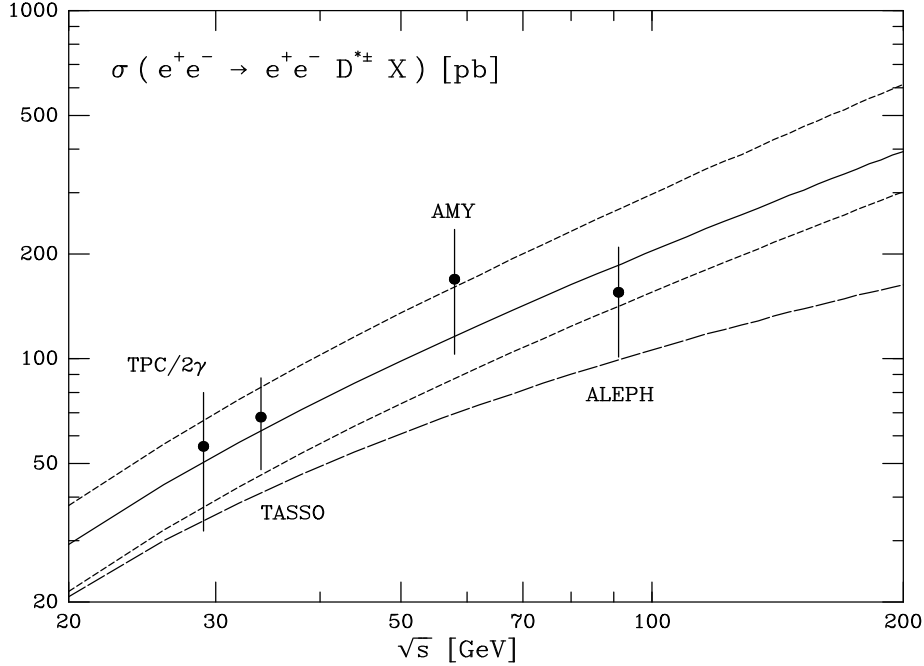,%
height=9.cm,%
width=12.5cm,%
bbllx=1.0cm,%
bblly=1.9cm,%
bburx=19.4cm,%
bbury=26.7cm,%
rheight=7.5cm,%
rwidth=15cm,%
angle=-90}

\vspace*{1.1cm}

\caption[xx]{\label{f_exp}
                 Total cross section for $\epem\to\epem D^{*\pm} X$
                 as function of the \epem collider energy. Shown is
                 the NLO prediction for $\mu=m_c=1.3$~GeV (upper
                 short-dashed curve), $\mu=2\,m_c; m_c=1.8$~GeV
                 (lower short-dashed curve), $\mu=\sqrt{2}\,m_c; m_c=1.5$~GeV
                 (solid curve) and the prediction for the direct process
                 in leading-order, $\mu=\sqrt{2}\,m_c; m_c=1.5$~GeV,
                 (long-dashed curve). The results are
                 compared to the recent experimental data
                 \cite{ggdata}. The TPC/2$\gamma$ and
                 TASSO results have been adjusted according to the
                 latest values for the $D^{*+}$ and $D^0$ branching
                 ratios.}


\end{figure}

In Fig.~\ref{f_exp} the leading-order and next-to-leading-order calculations
for two-photon production of $D^*$ mesons are compared with experimental data
from recent measurements \cite{ggdata}.
The predictions of the cross sections appear to be theoretically firm. A
variation of the charm quark mass $m$ and the renormalization/factorization
scale $\mu$ in the range 1.3~GeV $<$ $\mu$ $<$ $2m_c$ and 1.3~GeV $<$ $m_c$
$<$ 1.8~GeV leads to a total theoretical uncertainty of $\pm$~30~\% at LEP
energies. Even though the statistical and systematic errors are large,
the measured charm cross sections appear to overshoot the values estimated
at the Born level in the direct channel consistently. Adding however the
resolved contributions and the QCD radiative corrections, the agreement with
the recent $D^*$ data from PETRA, PEP, TRISTAN and LEP \cite{ggdata}
is quite satisfactory. The TOPAZ collaboration reports an excess of $D^*$
data at high transverse momenta which still has to be understood.
The VENUS, TOPAZ and AMY collaborations have measured open charm production
by tagging leptons from the semi-leptonic charm quark decays \cite{ggdata}.
Their results are consistent with the theoretical expectation but seem to
favour a rather low charm quark mass and a strong contribution from the
resolved process.

\newpage

{\noindent\large\bf
III.~Heavy-Quark Production in Photon-Proton Collisions}\\[3mm]
The production of heavy quarks in photon-proton collisions has been
studied extensively in fixed target experiments. Total cross sections,
single-inclusive distributions as well as heavy-quark correlations have
been calculated to next-to-leading-order accuracy \cite{gpnlo}. All
recent total-cross-section data on photoproduction of charm quarks are
in reasonable agreement with the NLO QCD predictions, once the theoretical
uncertainties are properly taken into account. Supplementing the NLO
predictions with a simple parametrization of the most important
non-perturbative effects, such as heavy flavoured hadron formation and
a moderate intrinsic transverse momentum of the initial state partons, is
sufficient to reproduce the experimental data on single- and
double-differential distributions \cite{rid}.\footnote{This is in contrast
to the case of heavy quark hadroproduction where the experimental
distributions are much harder than the theoretical expectations \cite{rid}.}
The $ep$ collider HERA will provide us with information on the dynamics of
heavy flavour production in a kinematical range very different from that
available at fixed target experiments. Among the topics to be studied at
HERA are photoproduction of charm and bottom quarks in both direct and
resolved-photon processes as well as heavy flavour production in
deep-inelastic scattering \cite{epnlo}. Bottom production is a particular
interesting subject to be studied at HERA since the larger value of the
bottom quark mass makes QCD predictions more reliable.
First HERA results on the total
charm photoproduction cross section have been published recently \cite{zeus1}
and found to be well described by the NLO QCD predictions. The theoretical
calculations for the total cross section are based on the standard
factorization formula \cite{coll} where the heavy quark mass $m \gg
\Lambda_{\mbox{\scriptsize QCD}}$ is considered as the large scale. In
next-to-leading order potentially large terms $\sim \alpha_s\ln(p_\perp/m)$
arise from collinear emission of gluons by a heavy quark at large transverse
momentum or from almost collinear branching of gluons or photons into heavy
quark pairs. These terms are not expected to affect the total production
rates, but they might spoil the convergence of the perturbation series at
$p_\perp\gg m$. An alternative way of making predictions at large $p_\perp$
is to treat the heavy quarks as massless partons. The mass singularities of
the form $\ln(p_\perp/m)$ are then absorbed into structure and fragmentation
functions in the same way as for the light $u,d,s$ quarks. This approach has
been used in Ref.~\cite{matteo} to study the production of large $p_\perp$
hadrons containing bottom quarks in $p\bar{p}$ collisions. Of course, in the
massless scheme the heavy quark is considered to be one of the massless
active flavours in the proton and, for the resolved contribution, of the
photon structure function. In Ref.~\cite{kkks} the differential distributions
for charm quark production at HERA have been calculated in the massive and
the massless approaches up to NLO accuracy.\footnote{A similar study, though
in leading-order only, has been presented by M.~Drees and R.~Godbole at this
meeting \cite{drgod}.}
The cross section in the massless scheme is approximately $70-100\%$ larger
than in the massive scheme. This difference can be attributed to the
contribution of the charm quark in the photon (for a more detailed discussion
and comparison between massive and massless schemes see Ref.~\cite{kkks}). A
measurement of heavy quark production at large $p_\perp$ at HERA will provide
information about the heavy flavour content of the proton and in particular
the photon. Such measurements will be very instructive since theoretical
opinions on that issue are rather divided \cite{intrin}.

The measurement of the gluon distribution in the nucleon is one of the
important goals of lepton-nucleon scattering experiments. The classical
methods exploit the evolution of the nucleon structure functions with
the momentum transfer and the size of the longitudinal structure function.
At HERA energies, however, the production of heavy quark states becomes
an important complementary tool. Besides open charm and
bottom production, the formation of $J/\psi$ bound states in inelastic
photoproduction experiments provides an experimentally attractive method
since $J/\psi$ particles are easy to tag in the leptonic decay modes.
Inelastic $J/\psi$ photoproduction through photon-gluon fusion is described
in the colour-singlet model through the subprocess
$\gamma + g \to J/\psi + g $ \cite{bj81}. Colour conservation and the
Landau-Yang theorem require the emission of a gluon in the final state.
The cross section is generally calculated in the static approximation in
which the motion of the charm quarks in the bound state is neglected. In this
approximation the production amplitude factorizes into the short distance
amplitude $\gamma + g \to c\overline{c} + g$, with $c\overline{c}$ in the
colour-singlet state and zero relative velocity of the quarks, and the
$c\overline{c}$ wave function $\varphi(0)$ of the $J/\psi$ bound state at the
origin which is related to the leptonic width. Relativistic corrections due
to the motion of the charm quarks in the $J/\psi$ bound state have been
demonstrated to be small in the inelastic region \cite{jkgw93}.
The calculation of the higher-order perturbative QCD corrections has been
performed recently \cite{phd,kzsz,mk}. Inclusion of the NLO corrections
reduces the scale dependence of the theoretical prediction considerably and
increases the cross section in the energy range of the fixed target
experiments, $E_\gamma \sim 100$~GeV, by more than 50\%. A comparison of the
next-to-leading order calculation with data from the two fixed-target
photoproduction experiments \cite{jpexp} reveals that the $J/\psi$ energy
dependence d$\sigma/\mbox{d}z$ ($z$ is the scaled $J/\psi$ energy,
$z=E_{J/\psi}/E_\gamma$) is adequately accounted for by the colour-singlet
model in the inelastic region $z<0.9$. Thus the shape of the
gluon distribution in the nucleon can be extracted from $J/\psi$
photoproduction data with confidence. Similar to the case of open heavy
flavour production, the absolute normalization of the cross section shows
a strong dependence on the value of the charm quark mass. In the static
approximation the choice $m_c = M_{J/\psi}/2$ is required for a consistent
description of the heavy bound state formation. However, a smaller mass value
might be appropriate for a reasonable description of the charm quark creation
in the hard scattering process. Taking $m_c = 1.4$~GeV and allowing for
higher-twist uncertainties of order $(\Lambda/m_c)^k \;\simlt\; 30\%$ for
$k \ge 1$, one can conclude that the normalization too appears to be under
semi-quantitative control \cite{mk}.

\begin{figure}[hbtp]

\hspace*{1.75cm}
\epsfig{%
file=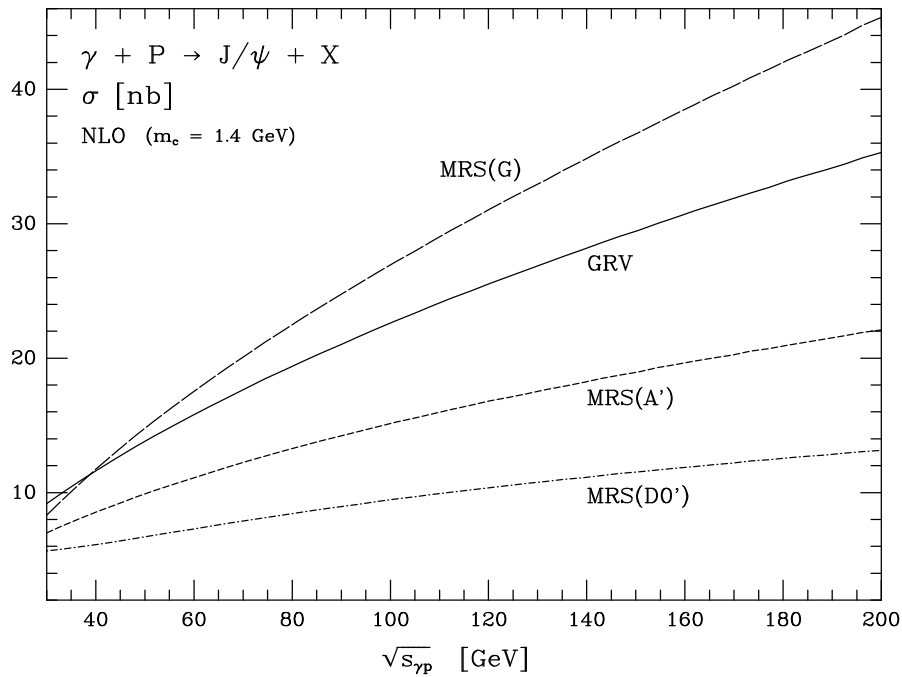,%
height=9.cm,%
width=12.5cm,%
bbllx=1.0cm,%
bblly=1.9cm,%
bburx=19.4cm,%
bbury=26.7cm,%
rheight=7.5cm,%
rwidth=15cm,%
angle=-90}

\vspace*{1.1cm}

\caption[xx]{
    \label{f_gptot}
    Total cross section for inelastic $J/\psi$ photoproduction
    $\gamma + P \to  J/\psi + X$ as a function of the
    photon-proton center of mass energy in the HERA energy range
    for different parametrizations of the gluon distribution of
    the proton \cite{jpglue}.}


\end{figure}

In Fig.\ref{f_gptot} the NLO prediction of the cross section is presented
for the \mbox{HERA} energy range.
Since the momentum fraction of the partons at \mbox{HERA} energies is
small, the cross section is sensitive to the parametrization of the gluon
distribution in the small-$x$ region $<\! \xi\! > \sim 0.001$.
As the absolute normalization of the cross section is rather sensitive to the
value of $\alpha_s$ and the charm quark mass, the discrimination between
different parametrizations of the gluon density in the proton has to rely on
the shape of the cross section (as a function of the photon-proton center of
mass energy) rather than the absolute size of the prediction.
First results on inelastic $J/\psi$ photoproduction at HERA have been
presented at this conference \cite{karshon}. It seems that the theoretical
expectation, Fig.~\ref{f_gptot}, underestimates the absolute normalization of
the data. With the present statistics, however, no firm conclusion can be
drawn.

The cross sections for the inelastic photoproduction of $\psi'$ particles
are expected to be 1/4 of the corresponding $J/\psi$ cross sections. The
production of $\Upsilon$ bottonium bound states is suppressed, compared with
$J/\psi$ states, by a factor of about 300 at HERA, a consequence of the
smaller bottom electric charge and the phase space reduction by the large
$b$ mass. The $P$-wave charmonium states $\chi_{c1}$ and $\chi_{c2}$ can be
detected in the radiative decay modes $\chi \to J/\psi + \gamma$ at HERA
\cite{kkz}. The predictions for the HERA energy range provide a crucial test
for the underlying picture as developed so far in the perturbative QCD sector.

{\noindent\bf Acknowledgements}\\
Thanks to Fred Combley, David Miller and their colleagues for
organizing an interesting and enjoyable meeting.



\begin{thebibliography}{99}

\bibitem{coll} J.C.~Collins, D.E.~Soper, G.~Sterman, \np{263}{86}{37}.

\bibitem{phsplit} T.F.~Walsh, P.M.~Zerwas, \pl{44}{73}{195};
                  E.~Witten, \np{120}{77}{189}.

\bibitem{photstruc} M.~Drees, K.~Grassie, \zp{28}{85}{451};
                    M.~Gl\"uck, E.~Reya, A.~Vogt, \prd{46}{92}{1973}.
                    L.E.~Gordon, J.K.~Storrow, \zp{56}{92}{307} and these
                    proceedings; P.~Aurenche, J.P.~Guillet, M.~Fontannaz,
                    preprint ENSLAPP-A-435-93-REV (1994); G.A.~Schuler,
                    T.~Sj\"ostrand, preprint CERN-TH/95-62 (1995); H.~Kan,
                    these proceedings.

\bibitem{eric} E.~Laenen, S.~Riemersma, preprint CERN-TH/95-103 (1995);
               E.~Laenen, S.~Riemersma, J.~Smith, W.L.~van~Neerven,
               \prd{49}{94}{5753}.

\bibitem{phd}  M.~Kr\"amer, PhD.~Thesis, Univ.~of~Mainz, 1994.

\bibitem{dkzz} M.~Drees, M.~Kr\"amer, J.~Zunft, P.M.~Zerwas,
               \pl{306}{93}{371}.

\bibitem{ggdata} W.~Braunschweig et al.\ [TASSO], \zp{47}{90}{499};
                 M.~Alston-Garnjost et al.\ [TPC/2$\gamma$],\pl{252}{90}{499};
                 R.~Enomoto et al.\ [TOPAZ], \prd{50}{94}{1879},
                 \pl{328}{94}{535}, M.~Iwasaki for the TOPAZ coll., these
                 proceedings; S.~Uehara et al.\ [VENUS], \zp{63}{94}{213};
                 T.~Aso et al.\ [AMY], preprint KEK-95-19,
                 T.~Nozaki for the AMY coll., these proceedings;
                 D.~Buskulic et al.\ [ALEPH], preprint CERN-PPE-95-40,
                 F.~Foster for the ALEPH coll., these proceedings.

\bibitem{gpnlo} P.~Nason, S.~Dawson, R.K.~Ellis, \np{303}{88}{607};
                G.~Altarelli, M.\ Diemoz, G.\ Martinelli, P.\ Nason,
                \np{308}{88}{724};
                W.~Beenakker, H.~Kuijf, W.~van~Neerven, J.~Smith,
                \prd{40}{89}{54};
                R.K.\ Ellis, P.\ Nason, \np{312}{89}{551};
                M.L.~Mangano, P.~Nason, G.~Ridolfi, \np{373}{92}{295}.
                J.~Smith, W.L.~van Neerven, \np{374}{92}{36};
                S.~Frixione, M.L.~Mangano, P.~Nason, G.~Ridolfi,
                \np{412}{94}{225}.

\bibitem{rid} G.~Ridolfi, S.~Frixione, M.L.~Mangano, P.~Nason,
              \np{431}{94}{453}; G.~Schuler, preprint CERN-TH/95-75 (1995).

\bibitem{epnlo} E.~Laenen, S.~Riemersma, J.~Smith, W.L.~van~Neerven,
                \np{392}{93}{162}, ibid.~229; B.W.~Harris, J.~Smith,
                preprint ITP-SB-94-06 and preprint ITP-SB-95-08 (1995).

\bibitem{zeus1} M.~Derrick et al. [ZEUS], \pl{349}{95}{225}.

\bibitem{matteo} M.~Cacciari, M.~Greco, \np{421}{94}{530}.

\bibitem{kkks}  B.A.~Kniehl, M.~Kr\"amer, G.~Kramer, M.~Spira, DESY preprint
                DESY-95-098 (1995).

\bibitem{drgod} M.~Drees and R.M.~Godbole, preprint LNF-95/020 (P) (1995).

\bibitem{intrin} M.~Gl\"uck, E.~Reya, M.~Stratmann, \np{422}{94}{37};
                 F.I.~Olness, S.T.~Riemersma, preprint SMU-HEP/94/21 (1994).

\bibitem{bj81} E.L.~Berger, D.~Jones, \prd{23}{81}{1521};
               R.~Baier, R.~R\"uckl, \pl{102}{81}{364};
               A.D.~Martin, C.-K.~Ng, W.J.~Stirling, \pl{191}{87}{200}.

\bibitem{jkgw93} H.~Jung, D.~Kr\"ucker, C.~Greub, D.~Wyler, \zp{60}{93}{721}.

\bibitem{kzsz} M.~Kr\"amer, J.~Zunft, J.~Steegborn, P.M.~Zerwas,
                 \pl{348}{95}{657}.

\bibitem{mk} M.~Kr\"amer, preprint DESY-95-046.

\bibitem{jpexp} R.\ Barate et al.\ [NA-14], \zp{33}{87}{505};
                B.H.\ Denby et al.\ [FTPS], \prl{52}{84}{795}.

\bibitem{jpglue} M.\ Gl\"uck, E.\ Reya, A.\ Vogt, preprint DO-TH 94/24;
                 A.D.\ Martin, R.G.\ Roberts, W.J.\ Stirling,
                 \pl{306}{93}{145} and preprint RAL-95-021.

\bibitem{karshon} U.~Karshon for the ZEUS coll., these proceedings.

\bibitem{kkz} W.~Kilian, M.~Kr\"amer, P.M.~Zerwas, DESY preprint in
              preparation.


\end{thebibliography}
\end{document}